\documentclass[12pt,a4paper]{article}
\usepackage{amsmath,amssymb,bm,ascmac}
\usepackage[dvipdfmx]{graphicx}
\usepackage{color}
\usepackage{authblk}
\begin{document}

\title{First-class constraints and the BV formalism}
\author{Ken KIKUCHI}
\affil{Department of Physics, Nagoya University}
\date{}
\maketitle

\makeatletter
\renewcommand{\theequation}
{\arabic{section}.\arabic{equation}}
\@addtoreset{equation}{section}
\makeatother

\newcommand{\defi}{\stackrel{\mathrm{def}}{\iff}}
\newcommand{\tdif}[2]{\frac{d#1}{d#2}}
\newcommand{\pdif}[2]{\frac{\partial #1}{\partial #2}}
\newcommand{\ddif}[2]{\frac{\delta #1}{\delta #2}}
\newcommand{\inj}{\hookrightarrow}
\newcommand{\surj}{\twoheadrightarrow}
\newcommand{\map}[1]{\stackrel{#1}{\rightarrow}}
\newcommand{\longmap}[1]{\stackrel{#1}{\longrightarrow}}
\newcommand{\dom}[1]{\mathrm{dom}\left(#1\right)}
\newcommand{\cod}[1]{\mathrm{cod}\left(#1\right)}
\newcommand{\notsubset}{\not\subset}
\newcommand{\notsupset}{\not\supset}
\newcommand{\fsl}[1]{\not\!#1}
\newcommand{\ld}[1]{\mathfrak L_{#1}}
\newcommand{\dk}[2]{\frac{d^{#1}#2}{i(2\pi)^{#1}}}
\newcommand{\dz}[1]{\frac{d#1}{2\pi i}}
\newcommand{\dv}[2]{d^D#1\sqrt{-#2}}
\newcommand{\delB}{\bm\delta_\mathrm{B}}
\newcommand{\E}{_\mathrm{E}}

\newcommand{\nn}{\nonumber}
\newcommand{\nabb}{\bm{\nabla}}
\newcommand{\mbb}{\mathbb}
\newcommand{\mcal}{\mathcal}
\newcommand{\mfra}{\mathfrak}
\newcommand{\subd}{_{(d)}}
\newcommand{\mh}{\hat{\mu}}
\newcommand{\nh}{\hat{\nu}}
\newcommand{\rh}{\hat{\rho}}
\newcommand{\sh}{\hat{\sigma}}
\newcommand{\mn}{\mu\nu}
\newcommand{\rs}{\rho\sigma}
\newcommand{\p}{(\phi)}
\newcommand{\hp}{(\hat\phi)}
\newcommand{\vp}{\varphi}
\newcommand{\dd}{\ddagger}
\newcommand{\mL}{\mathcal L}
\newcommand{\dl}[2]{\frac{\delta^L#1}{\delta#2}}
\newcommand{\dr}[2]{\frac{\delta^R#1}{\delta#2}}
\newcommand{\loc}{_\text{loc}}
\newcommand{\locw}[1]{_{\text{loc};#1-d}}
\newcommand{\Kg}{K_\gamma}
\newcommand{\Kp}{K_\phi}
\newcommand{\KA}{K_A}
\newcommand{\Ko}{K_\omega}
\newcommand{\Kbo}{K_{\bar\omega}}
\newcommand{\KB}{K_B}
\newcommand{\Kc}{K_c}
\newcommand{\Kbc}{K_{\bar c}}
\newcommand{\Kb}{K_b}
\newcommand{\Rg}{R_\gamma}
\newcommand{\Rp}{R_\phi}
\newcommand{\RA}{R_A}
\newcommand{\Ro}{R_\omega}
\newcommand{\Rbo}{R_{\bar\omega}}
\newcommand{\RB}{R_B}
\newcommand{\Rc}{R_c}
\newcommand{\Rbc}{R_{\bar c}}
\newcommand{\Rb}{R_b}

\begin{abstract}
Employing the Batalin-Vilkovisky (BV) formalism, we present a systematic and simple prescription to derive (first-class) constraints including the Hamiltonian constraint (a.k.a. flow equation), which plays pivotal role in holographic computation of Weyl anomalies. In this method, you do not have to compute canonical momenta nor Hamiltonians. Thus it may equip us with a `Lagrangian treatment' of constrained systems. We also point out an interesting analogy between antifields and first-class constraints.
\end{abstract}

\tableofcontents

\newpage
\setcounter{section}{+0}
\setcounter{subsection}{+0}
\section{Introduction}
The late Steve Jobs emphasized the importance of ``connecting the dots" \cite{Jobs}. Such connections (or analogies) of seemingly different areas also have been helped human beings to reveal new faces of physics (or furthermore, its relation to mathematics). A remarkable example would be a mysterious connection between statistical physics and quantum field theories \cite{Polyakov}. The analogy led us to important concepts such as renormalization group, or spontaneous symmetry breaking. In this paper, I would like to pursue another.

The analogy is a similarity of the flow equation (a.k.a. the Hamiltonian constraint) and the antibracket. The former has a form $\{S,S\}$ \cite{BVV,FMS00,FMS03}, and the latter has a form $(S,S)$ \cite{BV1,BV2,BV3}. This resemblance is not just a superficial one. Both of them are second order differential equations and they both encode information of constraints. As you might expect, the latter is more general, and the use of the BV formalism enables us to derive (first-class) constraints systematically. The prescription we present is so simple that it can be expressed in a single equation (\ref{const}).

The structure of this paper is as follows: In section \ref{BV}, we briefly review the BV formalism and explain the general procedure to derive (first-class) constraints. After some warm-up in section \ref{exercise}, we tackle the problem we want to deal with, i.e., the derivation of (first-class) constraints including the Hamiltonian constraint in section \ref{flow}. We conclude with some comments in section \ref{summary}.

\section{The BV formalism revisited}\label{BV}
In the BV formalism \cite{BV1,BV2,BV3}, one assembles fields into a collection simply called fields, e.g. $\Phi^n=(\gamma,\phi,A,c,\bar c)$, and for each component of the fields one introduces external field called antifield, e.g. $K_n=(\Kg,\Kp,\KA,\Kc,\Kbc)$. Then for arbitrary functionals $F[\Phi,K]$ and $G[\Phi,K]$, a bracket called the antibracket is defined by
\begin{equation}
	(F,G):=\int d^DX\Bigg\{\dr F{\Phi^n(X)}\dl G{K_n(X)}-\dr F{K_n(X)}\dl G{\Phi^n(X)}\Bigg\},\label{antibra}
\end{equation}
where summation over $n$ is understood, and the superscripts $R$ and $L$ indicates right and left derivatives, respectively. From the definition, we have
\[ (\Phi^m(X),K_n(Y))=\delta^m_n\delta^{(D)}(X-Y), \]
and antifields can be recognized as canonical momenta conjugate to fields in terms of antibrackets. This observation would be helpful later to require antifields be tensor densities.

The antibracket has following properties:
\[ (F,G)=(-)^{(\epsilon[F]+1)(\epsilon[G]+1)}(G,F),~~(-)^{\epsilon[F]\epsilon[H]+\epsilon[G]}(F,(G,H))+(\text{cyclic terms})=0 \]
for arbitrary functionals $F[\Phi,K]$, $G[\Phi,K]$, and $H[\Phi,K]$. Here, $\epsilon[F]$ denotes the statistics of a functional $F$ mod $2$. An extended action $S[\Phi,K]$ is defined as a solution of the (classical) master equation
\begin{equation}
	(S,S)=0.
\label{meq}
\end{equation}
Since the equation is a second order differential equation, we need two boundary conditions
\begin{equation}
	S[\Phi,K=0]=S_c[\varphi],\quad-\dr{S[\Phi,K]}{K_n}\Bigg|_{K=0}=R^n[\varphi,C]\label{bc}
\end{equation}
to fix a solution. Here I called original fields in a classical action $\varphi$ collectively, and `ghost fields' introduced in the BV formalism $C$ collectively. Therefore, a general solution (we will discuss further generalization later) is given by
\begin{equation}
	S[\Phi,K]\equiv S_c[\varphi]+S_K[\Phi,K],\label{sol}
\end{equation}
where one can show that the source term $S_K$ is linear in antifields $K$ (if algebras close off-shell \cite{Anselmi}\footnote{I mainly follow the convention of the paper.}):
\begin{equation}
	S_K[\Phi,K]:=-\int d^DXR^n[\Phi]K_n,\label{S_K}
\end{equation}
where summation over $n$ is understood. $R^n$ is nothing but a variation of $\Phi^n$. This can be easily seen if one considers an antibracket $(S,\Phi^n)$:
\begin{align*}
	(S[\Phi,K],\Phi^n(X))&=-\int d^DY\dr{S[\Phi,K]}{K_m(Y)}\dl{\Phi^n(X)}{\Phi^m(Y)}\\
	&=R^n[\Phi(X)].
\end{align*}

When one tries to quantize a theory $S_c$ with gauge symmetries, one has to fix a gauge. Since some tools will turn out to be useful later, we briefly review gauge fixing here, although we do not fix a gauge because we are interested in classical theories.

Let us define two spaces: $\mfra F:=\{\Phi,K\};$ spaces of (anti)field configurations and $\mcal F:=\{X:\mfra F\to\mbb K\};$ spaces of functionals over (anti)fields which have values in some field $\mbb K$. Since three natural operations, namely, scalar multiplication over $\mbb K$, addition, and product are defined in $\mcal F$, $(\mcal F,\stackrel{\mbb K}\cdot,+,\cdot)$ is an algebra\footnote{$\stackrel{\mbb K}\cdot$ denotes my personal notation for scalar multiplication over a field $\mbb K$.} with a unit element $1_{\mcal F}$. Then using some properties of the antibracket, one can show
\[ \forall X[\Phi,K]\in\mcal F,\ (S,(S,X))=0. \]
This means a map which is defined by an antibracket with the first entry chosen as the extended action
\[ (S[\Phi,K],\cdot):\mcal F\to\mcal F, \]
is nilpotent. In this sense, the BV formalism is said to generalize the BRST prescription. Being nilpotent, one can define a cohomology algebra on $\mcal F$:
\begin{align*}
	\mcal C:=&\{X\in\mcal F|(S,X)=0\},\\
	\mcal E:=&\{X\in\mcal F|\exists Y\in\mcal F\ s.t.\ X=(S,Y)\},
\end{align*}
where elements of these sets are called closed and exact, respectively, as usual. By defining an equivalence relation
\[ X,X'\in\mcal F\ ,\quad X\sim X'\defi X-X'\in\mcal E, \]
the quotient space $\mcal H:=\mcal C/\mcal E$ defines the cohomology algebra $(\mcal H,\stackrel{\mbb K}\cdot,+,\cdot)$\footnote{It is easy to see that the space is equipped with a well-defined product (defined as the one of ordinary product of functions). In fact, $\forall[F],[G]\in\mcal H$, consider a product of two equivalent classes $[F]$ and $[G]$, which is defined as an equivalent class of the products $F\cdot G$. Since $\mcal H$ is a quotient space, its elements are defined up to $\mcal E$, i.e., $F\sim F+(S,f)$, for some $f\in\mcal F$. Then
\begin{align*}
	[F+(S,f)]\cdot[G]&\equiv[F+(S,f)\cdot G]\\
	&=[F]\cdot[G]+[(S,f)\cdot G]\\
	&=[F]\cdot[G]+[(S,f\cdot G)]-(-)^{\epsilon[f]}[f(S,G)]=[F]\cdot[G].
\end{align*}
Moreover, since fields are assigned with various additive numbers such as ghost numbers or statistics, it is the graded cohomology algebra. That is, if we write subspaces with ghost number $n$ as $\mcal F_{(n)}$, $\mcal C_{(n)}$, and $\mcal E_{(n)}$, these subspaces do not have common parts, i.e.,
\[ \mcal F=\bigoplus_n\mcal F_{(n)},~~\mcal C=\bigoplus_n\mcal C_{(n)},~~\mcal E=\bigoplus_n\mcal E_{(n)}. \]
Therefore, the cohomology algebra decomposes to subspaces with fixed ghost numbers:
\[ \mcal H=\bigoplus_n\mcal H_{(n)}. \]
Note that since these functional are further assigned with statistics, they are also $\mbb Z_2$ graded, and these spaces can further be decomposed into Grassmann even and odd parts (although some of them may be trivially empty).\label{graded}}. Then the gauge fixing is realized by adding a cohomologically exact term $(S,\Psi)$ to the action
\begin{equation}
	S[\Phi,K]:=S_c[\Phi]+S_K[\Phi,K]+(S,\Psi)\equiv\int d^DX\mcal L(\Phi,K).\label{Sext}
\end{equation}
The functional $\Psi\in\mcal F$ has to be Grassmann odd to ensure $(S,\Psi)$ be Grassmann even, and that is the reason why $\Psi$ is called a gauge fermion. Furthermore, so as to justify adding the term to an extended action, $(S,\Psi)$ has to belong to the Lorentz singlet. This requirement can be stated in terms of condition on the gauge fermion $\Psi$. We will come back to the condition in short when we consider a gravitational theory.

The nilpotency of the BV transformation $(S,\cdot)$ guarantees invariance of the modified action $S'=S+(S,\Psi)$ under the transformation. Note that since the original extended action $S$ was defined as a solution of the master equation $(S,S)=0$, the modified action is also a solution of the master equation.

It is well-known that under a diffeomorphism transformation $X^M\mapsto X^{'M}(X):=
X^M+\xi^M(X)$ metrics transform as
\[ \delta\gamma_{MN}(X)=-\nabla_M\xi_N(X)-\nabla_N\xi_M(X). \]
Therefore, BRST transformations of the metric and corresponding ghosts are fixed by imposing the nilpotency $\delB^2=0$:
\begin{align*}
	\delB\gamma_{MN}(X)&=-\gamma_{LN}(X)\nabla_M\omega^L(X)-\gamma_{ML}(X)
		\nabla_N\omega^L(X),\\
	\delB\omega^M(X)&=-\omega^N(X)\nabla_N\omega^M(X),\\
	\delB\bar\omega_M(X)&=iB_M(X),\\
	\delB B_M(X)&=0.
\end{align*}
The BRST transformations of gauge fields and matter fields\footnote{We just consider scalar fields here. Generalizations to other matter fields would be straightforward.} in some representation $r$ of a gauge group $G$, whose representation matrices are denoted by $T^a$, are given as usual:
\begin{align*}
	\delB A_M(X)&=\nabb_Mc(X),\\
	\delB \phi^I(X)&=ig\Big(c(X)\phi(X)\Big)^I,\\
	\delB c(X)&=igc^2(X),\\
	\delB \bar c(X)&=ib(X),\\
	\delB b(X)&=0,
\end{align*}
or in components
\begin{align*}
	\delB A^a_M(X)&=\nabb_Mc^a(X),\\
	\delB \phi^I(X)&=gc^a(X)\Big(iT^a\phi(X)\Big)^I,\\
	\delB c^a(X)&=-\frac12gf^{abc}c^b(X)c^c(X),\\
	\delB \bar c^a(X)&=ib^a(X),\\
	\delB b^a(X)&=0.
\end{align*}
Here the covariant derivative $\nabb_M$ is defined to include not only the Levi-Civita connection $\Gamma$ but also the gauge connection $A$:
\[ \nabb_M{f^I_{N\cdots}}^{K\cdots}:=\nabla_M{f^I_{N\cdots}}^{K\cdots}-A^a_M\left(iT^a{f_{N\cdots}}^{K\cdots}\right)^I. \]

To get a grip of the prescription, let us consider a familiar example, the YM theory. As in the case of the BRST quantization, a gauge fermion
\[ \Psi[\Phi]:=\int d^DX\Bigg\{-\bar c^a\Big(\frac\alpha2ib^a+\mathcal G^a[\Phi]\Big)\Bigg\} \]
produces ghost terms and gauge fixing terms, where 
$\mathcal G^a$ is a gauge fixing functional.
In fact,
\begin{align*}
	(S,\Psi)&=-\int d^DXR^n[\Phi(X)]\dl\Psi{\Phi^n(X)}\\
	&=\int d^DX\Bigg\{\frac\alpha2b^ab^a-ib^a\mcal G^a[\Phi]+\bar c^aR^n[\Phi]\dl{\mcal G^a[\Phi]}{\Phi^n(X)}\Bigg\}.
\end{align*}
For instance, if one chooses the Lorentz gauge condition $\mcal G^a=\partial^MA^a_M$, the last term reduces to $-\int d^DX\partial^M\bar c^a\nabb_Mc^a$, which has the correct familiar form.

We now give some comments on representations. When a theory includes metric $\gamma$ in fields $\Phi^n$, we have to pay attention on powers of $\sqrt\gamma$. It is natural to require the BV transformation $(S,\cdot)$ preserve representations of the Lorentz group because we expect fields $\Phi^n$ and their variations $R^n$ to belong to the same representations. Then, to justify adding $(S,\Psi)$ to an extended action, $\Psi$ itself must belong to a Lorentz singlet. This is the condition on the gauge fermion as informed before. On the other hand, the source term $S_K$ of the extended action must also belong to a Lorentz singlet. This observation and the definition (\ref{S_K}) lead us to a conclusion that when a theory includes metrics, antifields must be tensor densities. The result is similar to the fact that canonical momenta in the Hamiltonian formulation of the general relativity are tensor densities \cite{Wald}. This result also supports our observation to recognize antifields as canonical momenta conjugate to fields. Since we required the variations $R^n$ belong to the same representation of the Lorentz group as the original fields $\Phi^n$, we can simply employ the familiar expression of the BRST variations for $R^n$; $R^n[\Phi]=\delB\Phi^n$. We have already used the fact above.

Equipped with the machinery, a procedure to obtain (first-class) constraints can be stated very concisely: just consider antibrackets $(S,K_n)$, and to reproduce the original theory, set all antifields to zero after the computation. Since we introduce antifields as external fields, the last procedure would be allowed. We give another justification later. As we explained above, a term of the form $(S,f)$ is an element of $\mcal E$, and is equivalent to $0\in\mcal H$. In other words, the term is cohomologically trivial $(S,f)\sim0$. Therefore our procedure to obtain (first-class) constraints can be schematically summarized as
\begin{equation}
	(S,K_n)\Big|_{K=0}\sim0\in\mcal H.\label{const}
\end{equation}
This is the whole story of our procedure, and you do not have to compute canonical momenta, nor Legendre transform to get Hamiltonians.

Finally, we mention another interesting similarity between antifleds and first-class constraints. That is, I would like to propose a viewpoint to recognize antifields as first-class constraints. This statement may start to look plausible if one notices the following observations. Since we introduce antifields as independent external fields in the sense that $(K_m,K_n)=0$, they commute with respect to the antibracket. Secondly, a linear combination of antifields, that is the source term $S_K$, serves as a generator of gauge transformations $(S_K,\Phi^n(X))=R^n[\Phi(X)]$ so does a linear combination of first-class constraints in the canonical treatment of constrained systems a la Dirac. Furthermore, the extented action is defined by adding a linear combination of antifields to the classical action (up to cohomologically trivial terms) as one defines extended (or total) Hamiltonian by adding a linear combination of constraints to the original Hamiltonian.

Once this claim is accepted, the manipulation to set antifields zero after computation may also start to look natural in analogy with the fact that first-class constraints are weakly zero. In addition, one immediately notices that if the BV transformation is recognized as `time evolution', consistency requires the BV transformation of constraints vanish. In our language, this requirement is equivalent to claim that the variation be cohomologically trivial. This requirement is automatically satisfied thanks to the nilpotency of the BV transformation, and these do not produce further constraints.

Therefore, we may be led to a `Lagrangian treatment' of constrained systems which bypasses computations of canonical momenta or Hamiltonians by employing the following identifications:
\begin{align*}
	q&\leftrightarrow\Phi;\text{generalized coordinates}\\
	p&\leftrightarrow K;\text{canonical momenta}\\
	\{\cdot,\cdot\}_P&\leftrightarrow(\cdot,\cdot);\text{brackets}\\
	H&\leftrightarrow S;\text{time evolution generator}\\
	\phi&\leftrightarrow K;\text{first-class constraints}.
\end{align*}

\section{Exercises with simple theories}\label{exercise}
In the context of the AdS/CFT correspondence, one usually studies a classical bulk action\footnote{In order to ensure the action be bounded from below, we assume $L_{IJ}\p$ be positive definite. As an immediate consequence, there exists its inverse $L^{-1}\p$.}
\begin{align}
	S_c&\left[\gamma_{MN},\phi^I,A_M\right]\nn\\
	=&\int_{M_D}d^DX\sqrt{\gamma}\,\Big\{V\p-R_{(D)}+\frac12L_{IJ}\p\gamma^{MN}\nabb_M\,\phi^I\nabb_N\,\phi^J+\frac14B\p F^a_{MN}F^{aMN}\Big\}.
\label{bulkS}
\end{align}
This is the action we would like to apply the BV formalism.

As stated, in the BV formalism, one introduces antifields $K_n$ for all components of the fields $\Phi^n=(\gamma,\phi,A,\omega,\bar\omega,B,c,\bar c,b)$:
\[ K_n=(K_\gamma,K_\phi,K_A,K_\omega,K_{\bar\omega},K_B,K_c,K_{\bar c},K_b). \]
These fields and antifields have statistics, ghost numbers, and the other quantum numbers as in the table below:
\begin{table}[ht]
\begin{center}
\begin{tabular}{c|c|c|c}
(anti)field&$\epsilon[\cdot]$ mod2&ghost \#&$G$ \\\hline\hline
$\gamma$&0&0&1\\\hline
$\phi$&0&0&$r$\\\hline
$A$&0&0&Adj\\\hline
$\omega$&1&1&1\\\hline
$\bar\omega$&1&-1&1\\\hline
$B$&0&0&1\\\hline
$c$&1&1&Adj\\\hline
$\bar c$&1&-1&Adj\\\hline
$b$&0&0&Adj\\\hline\hline
$\Kg$&1&-1&1\\\hline
$\Kp$&1&-1&$\bar r$\\\hline
$\KA$&1&-1&Adj\\\hline
$\Ko$&0&-2&1\\\hline
$\Kbo$&0&0&1\\\hline
$\KB$&1&-1&1\\\hline
$\Kc$&0&-2&Adj\\\hline
$\Kbc$&0&0&Adj\\\hline
$\Kb$&1&-1&Adj
\end{tabular}~~.
\end{center}
\caption{Assignment of (quantum) numbers}
\end{table}

Since the variations of fields have been given, it is straightforward to find an extended action:
\begin{equation}
	S[\Phi,K]=S_c[\Phi]+S_K[\Phi,K]\equiv\int d^DX\mcal L(\Phi,K)\label{S}
\end{equation}
where $S_c$ is nothing but (\ref{bulkS}) and the source term is given by
\begin{align}
	S_K[\Phi,K]:=&-\int d^DX\Big\{{\Rg}_{MN}\Kg^{MN}+\Rp^I{\Kp}_I+{\RA}^a_M\KA^{aM}+\Ro^M{\Ko}_M\nn\\
	&~~~~~~~~+{\Rbo}_M\Kbo^M+{\RB}_M\KB^M+\Rc^a\Kc^a+\Rbc^a\Kbc^a+\Rb^a\Kb^a\Big\}\label{SK}\\
	=&-\int d^DX
\Big\{(\delB\gamma)_{MN}\Kg^{MN}+(\delB\phi)^I{\Kp}_I+(\delB A)_M^a\KA^{aM}+(\delB\omega)^M{\Ko}_M\nn\\
	&~~~~~~~~+(\delB\bar\omega)_M\Kbo^M+(\delB B)_M\KB^M+(\delB c)^a\Kc^a+(\delB\bar c)^a\Kbc^a+(\delB b)^a\Kb^a\Big\}.\nn
\end{align}
Recall that we do not fix a gauge because we are just interested in classical theories.


Now that we are ready to face the flow equation (a.k.a. Hamiltonian constraint). However, since the problem is a little complicated, it would be instructive to exercise with simpler theories. The analysis will also help to make us confident that we are in the right direction.

\subsection{Example 1 : (slightly modified) scalar QED}\label{sqed}
Let us begin with the well-studied example, the scalar QED (though slightly modified kinetic term of scalar fields for the later convenience)\footnote{Since we want to treat all theories in this paper classically, we do not distinguish bare quantities and renormalized ones.}:
\[ S_c[A,\phi]=\int d^DX\Big\{\frac14F_{MN}F^{MN}+V\p+\frac12L_{IJ}\p D^M\phi^ID_M\phi^J\Big\}, \]
where $D_M\phi^I:=\partial_M\phi^I-A_M(ig\phi^I)$. $\Phi^n=(A,\phi,c,\bar c,b)$, $K_n=(\KA,\Kp,\Kc,\Kbc,\Kb)$ and the antibracket is defined by
\[ (F,G):=\int d^DX\Bigg\{\dr F{\Phi^n(X)}\dl G{K_n(X)}-\dr F{K_n(X)}\dl G{\Phi^n(X)}\Bigg\}. \]
The extended action is given by 
\[ S[\Phi,K]=S_c[\Phi]+S_K[\Phi,K] \]
where
\begin{align*}
	S_K[\Phi,K]&=-\int d^DX\Big\{{\RA}_M\KA^M+\Rp^I{\Kp}_I+\Rc\Kc+\Rbc\Kbc+\Rb\Kb\Big\}\\
	&=-\int d^DX\Big\{(D_Mc)\KA^M+(igc\phi^I){\Kp}_I+(ib)\Kbc\Big\}.
\end{align*}
It is known that the system has a first-class constraint known as the Gauss's law. To see if the BV formalism correctly reproduces the constraint, let us consider a candidate antibracket $(S,\KA^M)$:
\begin{align}
	(S,\KA^\tau(X))&=\int d^DY\dr S{\Phi^n(Y)}\dl{\KA^\tau(X)}{K_n(Y)}\nn\\
	&=\dr S{A_N(X)}\delta^\tau_N\nn\\
	&=\partial_KF^{\tau K}(X)-L_{IJ}\p(ig\phi^I)D^\tau\phi^J(X)\sim0.\label{Gauss sQED}
\end{align}
Our claim is that this is nothing but the Gauss's law. Note that we have not computed canonical momenta nor Hamiltonian.

Although (\ref{Gauss sQED}) does not seem the same as the conventional form, it is also possible to see a complete agreement if one rewrites (\ref{Gauss sQED}) in terms of Hamiltonian variables, however, I emphasize again that this is not necessary in our procedure.

Since\footnote{We also have
\[ \pi_c:=\pdif{^L\mcal L}{(\partial_\tau c)}=-\KA^\tau,~~\pi_{\bar c}:=\pdif{^L\mcal L}{(\partial_\tau\bar c)}=0,~~\pi_b:=\pdif{\mcal L}{(\partial_\tau b)}=0. \]}
\[ \pi^M:=\pdif{\mcal L}{(\partial_\tau A_M)}=F^{\tau M},~~\pi_I:=\pdif{\mcal L}{(\partial_\tau\phi^I)}=L_{IJ}\p D^\tau\phi^J, \]
in the language of canonical variables, the antibracket reduces to
\[ (S,\KA^\tau(X))=\partial_K\pi^K(X)-(ig\phi^I)\pi_I(X)
	\sim0\in\mcal H_{(0)}. \]
This is nothing but the Gauss's law.
We expect one can obtain dozens of constraints by computing antibrackets $(S,K_n)$ without mentioning conjugate momenta, and some of them are studied below.

\subsection{Example 2 : scalar coupled to gravity}\label{sgra}
The classical (bulk) action is now given by
\begin{align*}
	S_c[\gamma,\phi]=&\int_{M_D}d^DX\sqrt\gamma\,\Big\{V\p-R_{(D)}
+\frac12L_{IJ}\p\gamma^{MN}\partial_M\,\phi^I\partial_N\,\phi^J\Big\}.
\label{bulkS'}
\end{align*}
The fields are given by $\Phi^n=(\gamma,\phi,\omega,\bar\omega,B)$ and the corresponding antifields are $K_n=(\Kg,\Kp,\Ko,\Kbo,\KB)$. Then the extended action is given by
\[ S[\Phi,K]=S_c[\Phi]+S_K[\Phi,K]\equiv\int d^DX\mcal L(\Phi,K), \]
where the source term $S_K$ is defined by
\begin{align*}
	S_K[\Phi,K]:=&-\int d^DX\Big\{{\Rg}_{MN}\Kg^{MN}+\Rp^I{\Kp}_I+\Ro^M{\Ko}_M+
{\Rbo}_M\Kbo^M+
{\RB}_M\KB^M\Big\}\\
	=&-\int d^DX\Big\{(-\gamma_{LN}\nabla_M\omega^L-\gamma_{ML}\nabla_N\omega^L)\Kg^{MN}+(-\omega^L\partial_L\phi^I){\Kp}_I\\
	&~~~~~~~~~~~~~~~~~~~~+(-\omega^L\nabla_L\omega^M){\Ko}_M+(iB_M)\Kbo^M\Big\}.
\end{align*}
Note that, as mentioned in section \ref{BV}, we defined $\mcal L$ as a scalar density, namely, we include $\sqrt\gamma$. The BV transformation $(S,\cdot)$ of antifields yields, for example,
\begin{align*}
	(S,\Kg^{MN}(X))&=\dr S{\gamma_{MN}(X)}\\
	&=\dr{S_c}{\gamma_{MN}(X)}+\dr{S_K}{\gamma_{MN}(X)}
\end{align*}
where
\begin{align*}
	\dr{S_c}{\gamma_{MN}}&=\frac12\sqrt\gamma\gamma^{MN}\left\{V\p-R_{(D)}+\frac12L_{IJ}\p\gamma^{KL}(X)\partial_K\phi^I\partial_L\phi^J\right\}\\
	&~~~~+\sqrt\gamma R^{MN}-\frac12\sqrt\gamma L_{IJ}\p\partial^M\phi^I\partial^N\phi^J,
\\
	\dr{S_K}{\gamma_{MN}}&=\nabla_L\omega^M\Kg^{NL}+\nabla_L\omega^N\Kg^{ML}-\nabla_L(\omega^L\Kg^{MN})\\
	&~~~~-\frac12\gamma^{LM}\nabla_K(\omega^K\omega^N{\Ko}_L)-\frac12\gamma^{LN}\nabla_K(\omega^M\omega^K{\Ko}_L).
\end{align*}
Then we can study candidate antibrackets which are expected to reproduce the first-class constraints. Let us start to attack the simpler constraint, the Hamiltonian constraint. Taking the consideration of Appendix \ref{ADM} into account, a candidate antibracket is given by $(S,K_N)$:
\begin{align}
	(S,K_N)&\equiv(S,2N\Kg^{\tau\tau})\nn\\
	&=2(S,N)\Kg^{\tau\tau}+2N(S,\Kg^{\tau\tau})\nn\\
	&=\sqrt h\Bigg\{V\p-R_{(D-1)}+\frac12L_{IJ}\p h^{\mn}\partial_\mu\phi^I\partial_\nu\phi^J\nn\\
	&~~~~~~~~+K^2-K^{\mn}K_{\mn}-\frac1{2N^2}L_{IJ}\p(\partial_\tau\phi^I-\lambda^\mu\partial_\mu\phi^I)(\partial_\tau\phi^J-\lambda^\nu\partial_\nu\phi^J)\Bigg\}\nn\\
	&~~~~+(K\text{ terms}).\label{Hamilsgra}
\end{align}
This is our final result, however, in terms of canonical variables\footnote{We also have
\[ \pi_{\omega M}:=\pdif{^L\mcal L}{(\partial_\tau\omega^M)}=2\gamma_{MN}\Kg^{\tau N}-\omega^\tau{\Ko}_M,~~\pi_{\bar\omega}^M:=\pdif{^L\mcal L}{(\partial_\tau\bar\omega_M)}=0,~~\pi_B^M:=\pdif{\mcal L}{(\partial_\tau B_M)}=0, \]
where the extrinsic curvature is defined by
\[ K_{\mn}:=\frac1{2N}\Big(\partial_\tau h_{\mn}-\nabla_\mu\lambda_\nu-\nabla_\nu\lambda_\mu\Big). \]
}
\[ \pi^{\mn}:=\pdif{\mcal L}{(\partial_\tau h_{\mn})}=\sqrt h\Big(K^{\mn}-h^{\mn}K\Big),~~\pi_I:=\pdif{\mcal L}{(\partial_\tau\phi^I)}=\frac{\sqrt h}NL_{IJ}\Big(\partial_\tau\phi^J-\lambda^\mu\partial_\mu\phi^J\Big)+\omega^\tau{\Kp}_I, \]
(\ref{Hamilsgra}) reduces to
\begin{align*}
	(S,K_N)&=\sqrt h\Bigg\{V\p-R_{(D-1)}+\frac12L_{IJ}\p h^{\mn}\partial_\mu\phi^I\partial_\nu\phi^J\\
	&~~~~~~~~+\frac1h\left(\frac1{D-2}\pi^2-\pi^{\mn}\pi_{\mn}\right)-\frac1{2h}(L^{-1}\p)^{IJ}\pi_I\pi_J\Bigg\}+(K\text{ terms}).\\
\end{align*}
Our prescription (\ref{const}) again correctly reproduces the first-class constraint.

Similarly, a candidate antibracket which is expected to produce the momentum constraint is given by $(S,K_\lambda^\mu)$:
\begin{align}
	(S,K_\lambda^\mu)&\equiv(S,2\lambda^\mu\Kg^{\tau\tau}+2\Kg^{\mu\tau})\nn\\
	&=\sqrt h\Bigg\{\frac2{N}h^{\mn}(R_{\tau\nu}-\lambda^\rho R_{\rho\nu})-\frac1NL_{IJ}\p h^{\mn}(\partial_\tau\phi^I-\lambda^\rho\partial_\rho\phi^I)\partial_\nu\phi^J\Bigg\}+(K\text{ terms}),\label{momsgra}
\end{align}
and in terms of canonical variables
\[ (S,K_\lambda^\mu)=2\nabla_\nu\pi^{\mn}-h^{\mn}\pi_I\partial_\nu\phi^I+(K\text{ terms}),
\]
which completely agrees the conventional expression.

\section{Derivation of the flow equation}\label{flow}
With experience and confidence we have obtained from the above exercises, let us tackle our theory (\ref{bulkS}). It would be instructive to consider the easiest constraint, the Gauss's law, first. A candidate equation is obtained by considering an antibracket $(S,\KA^{a\tau})$:
\begin{equation}
	(S,\KA^{a\tau})=
\nabb_\mu\Big[N\sqrt hB\p F^{a\tau\mu}\Big]-\frac{\sqrt h}NL_{IJ}\p\Big(\nabb_\tau\phi^I-\lambda^\mu\nabb_\mu\phi^I\Big)(iT^a\phi)^J-f^{abc}c^b\KA^{c\tau}\sim0.\label{Gauss}
\end{equation}
This is our final result, and in terms of canonical variables
\begin{align}
	\pi^{\mn}:=&\pdif{\mcal L}{(\partial_\tau h_{\mn})}\nn\\
	=&\sqrt h\Big(K^{\mn}-h^{\mn}K\Big),\label{pimn}\\
	\pi_I:=&\pdif{\mcal L}{(\partial_\tau\phi^I)}\nn\\
	=&\frac{\sqrt h}NL_{IJ}\Big(\nabb_\tau\phi^J-\lambda^\mu\nabb_\mu\phi^J\Big)+\omega^\tau{\Kp}_I,\label{piI}\\
	\pi^{a\mu}:=&\pdif{\mcal L}{(\partial_\tau A^a_\mu)}\nn\\
	=&N\sqrt hBF^{a\tau\mu}=\frac{\sqrt h}NB\Big(h^{\mn}F^a_{\tau\nu}-\lambda^\nu h^{\rho\mu}F^a_{\nu\rho}\Big),\label{piamu}
\end{align}
this reduces to
\[ (S,\KA^{a\tau})=\nabb_\mu\pi^{a\mu}-\pi_I(iT^a\phi)^I+(K\text{ terms}), \]
which again completely agrees with the conventional expression \cite{KS}.

Now that it would be clear that the flow equation we have been searching for would appear in an antibracket $(S,K_N)$ where $K_N$ is a component of the antifield $\Kg$ corresponding to the lapse function $N$:
\begin{align}
	(S,K_N)&\equiv(S,2N\Kg^{\tau\tau})\nn\\
	&=\sqrt h\Bigg\{V\p-R_{(D-1)}+\frac12L_{IJ}\p h^{\mn}\nabb_\mu\phi^I\nabb_\nu\phi^J+\frac14B\p h^{\mu\rho}h^{\nu\sigma}F^a_{\mn}F^a_{\rs}\nn\\
	&~~~~~~~~+K^2-K^{\mn}K_{\mn}-\frac1{2N^2}L_{IJ}\p(\nabb_\tau\phi^I-\lambda^\mu\nabb_\mu\phi^I)(\nabb_\tau\phi^J-\lambda^\mu\nabb_\mu\phi^J)\nn\\
	&~~~~~~~~-\frac{B\p}{2N^2}h^{\mn}\Big(F^a_{\tau\mu}-\lambda^\rho F^a_{\rho\mu}\Big)\Big(F^a_{\tau\nu}-\lambda^\sigma F^a_{\sigma\nu}\Big)\Bigg\}+(K\text{ terms}).\label{Hamil}
\end{align}
We claim it is the flow equation (or the Hamiltonian constraint) in a little disguised appearance. If you doubt if it is really the conventional one, you can convince yourself by simply rewriting it in terms of canonical variables \cite{KS}:
\begin{align*}
	(S,K_N)&=\sqrt h\Bigg\{V\p-R_{(D-1)}+\frac12L_{IJ}\p h^{\mn}\nabb_\mu\phi^I\nabb_\nu\phi^J+\frac14B\p h^{\mu\rho}h^{\nu\sigma}F^a_{\mn}F^a_{\rs}\\
	&~~~~~~~~+\frac1h\left(\frac1{D-2}\pi^2-\pi^{\mn}\pi_{\mn}\right)-\frac1{2h}(L^{-1}\p)^{IJ}\pi_I\pi_J-\frac1{2hB\p}h_{\mn}\pi^{a\mu}\pi^{a\nu}\Bigg\}\\
	&~~~~~~~~+(K\text{ terms}).
\end{align*}

Finally, let us also check if the formalism correctly reproduce the momentum constraint. A candidate antibracket is $(S,K_\lambda^\mu)$:
\begin{align}
	(S,K_\lambda^\mu)&\equiv(S,2\lambda^\mu\Kg^{\tau\tau}+2\Kg^{\mu\tau})\nn\\
	&=\sqrt h\Bigg\{\frac2Nh^{\mn}(R_{\tau\nu}-\lambda^\rho R_{\rho\nu})-\frac1NL_{IJ}\p h^{\mn}(\nabb_\tau\phi^I-\lambda^\rho\nabb_\rho\phi^I)\nabb_\nu\phi^J\nn\\
	&~~~~~~~~-\frac{B\p}Nh^{\mn}h^{\rs}F^a_{\nu\sigma}(F^a_{\tau\rho}-\lambda^\alpha F^a_{\alpha\rho})\Bigg\}+(K\text{ terms}).\label{moment}
\end{align}
In terms of canonical variables this reduces to
\[ (S,K_\lambda^\mu)=2\nabla_\nu\pi^{\mn}-h^{\mn}\pi_I\nabb_\nu\phi^I-h^{\mn}F^a_{\nu\rho}\pi^{a\rho}+(K\text{ terms}).
\]
Again, this is in complete agreement with the conventional expression \cite{KS}.

\section{Summary}\label{summary}
We have seen that in all cases we have studied, our prescription (\ref{const}) correctly reproduces (first-class) constraints. Furthermore, once the prescription is accepted, one does not have to compute canonical momenta nor Hamiltonians. In this sence, the prescription may uncover a possibility of `Lagrangian approach' to constrained systems\footnote{One usually treats Yang-Mills theory, a typical example of constrained systems, in a following way; one computes canonical momenta $\pi^{aM}$ conjugate to the non-Abelian gauge field $A^a_M$ and performs Legendre transformation to obtain the Hamiltonian $H$. Then one encounters a so called primary constraint $\pi^{a\tau}\approx0$. We have to set $\phi_1^a:=\pi^{a\tau}$ zero, however, it is not sufficient to impose the condition just at a `time' slice, rather one also has to set its `velocity' zero so as to ensure the constraint remain satisfied on every `time' slice. If the `velocity' is not automatically zero, this produces another constraint so called secondary. In fact, the Gauss's law appears as a secondary constraint $\dot\phi_1^a\propto\{\phi_1^a,H\}_P\approx0$. However, it is also possible to get the Gauss's law as an equation of motion for $A^a_\tau$. I would like to thank Zohar Komargodski for pointing out the possibility that all constraints would be obtained as equations of motion.}. We have also pointed out an interesting analogy between antifields and first-class constraints.

Some comments are in order. Firstly, one may wonder what shows up if one puts other antifields in the prescription (\ref{const}). After some considerations, one would learn that they just produce equations of motion. In the canonical formalism, one has to $discover$ constraints by, for example, looking for variables without `time' derivatives, however, the fact mentioned above implies that (first-class) constraints can be obtained on an equal footing as the equations of motion. In particular, one does not have to look for constraints of the system heuristically. All you have to do is to put antifields in (\ref{const}) one after another. The systematic and exhaustive character of our prescription may reveal some constraints which are not known at present.

Secondly, I have mentioned the similarity between antifields and first-class constraints, and proposed identifications of some objects in section \ref{BV}. However, we essentially used the fact that the source term $S_K$ is linear in antifields in the identification. Thus it would not hold in cases algebras do not close off-shell. It is also interesting to see whether the identification holds even after quantization.

Thirdly, all constraints we have considered are limited to first-class constraints. A natural future direction is thus to consider systems with second-class constraints. This is related to the problem we have avoided, that is, the gauge-fixing, because it is known that first-class constraints can be turned into second-class ones by fixing gauges.

Finally, we have paid little attention on surface terms, that is, have dropped the Gibbons-Hawking term throughout the paper. This would not cause problems as long as we are considering compact spacetime $M_D$, however, the term becomes relevant if we would like to study non-compact ones. We would like to come back to these problems in the future.

\section*{Acknowledgement}
The author would like to thank Zohar Komargodski and Tadakatsu Sakai for many helpful comments. He is also grateful for the hospitality of the Weizmann Institute of Science where the final stage of this work was done.

\makeatletter
\renewcommand{\theequation}
{\Alph{section}.\arabic{subsection}.\arabic{equation}}
\@addtoreset{equation}{subsection}
\makeatother
\appendix
\section{ADM decomposition}\label{ADM}
Often, one decomposes $D$-dimensional metrics into one-dimensional part and the others:
\begin{align*}
	\gamma_{MN}&=\begin{pmatrix}N^2+\lambda^\rho\lambda_\rho&\lambda_\nu\\\lambda_\mu&h_{\mu\nu}\end{pmatrix},\\
	\gamma^{MN}&=\frac1{N^2}\begin{pmatrix}1&-\lambda^\mu\\-\lambda^\nu&N^2h^{\mu\nu}+\lambda^\mu\lambda^\nu\end{pmatrix}.
\end{align*}
This is called the ADM decomposition. Employing the decomposition, the metric part of the source term (\ref{SK}) can also be decomposed as following:
\begin{align*}
	S_K&\ni-\int d^DX(\delB\gamma_{MN})\Kg^{MN}\\
	&=-\int d^DX\Bigg\{(\delB N)2N\Kg^{\tau\tau}+(\delB\lambda_\mu)\Big[2\lambda^\mu\Kg^{\tau\tau}+2\Kg^{\mu\tau}\Big]+(\delB h_{\mn})\Big[\Kg^{\mn}-\lambda^\mu\lambda^\nu\Kg^{\tau\tau}\Big]\Bigg\}.
\end{align*}
Thus if one defines
\[ K_N:=2N\Kg^{\tau\tau},~~K_\lambda^\mu:=2\lambda^\mu\Kg^{\tau\tau}+2\Kg^{\mu\tau},~~K_h^{\mn}:=\Kg^{\mn}-\lambda^\mu\lambda^\nu\Kg^{\tau\tau}, \]
the term can be written
\[ -\int d^DX\Bigg\{(\delB N)K_N+(\delB\lambda_\mu)K_\lambda^\mu+(\delB h_{\mn})K_h^{\mn}\Bigg\}. \]
This expression seems nice and the fact that the new antifields are also linear in the old antifields are satisfying.

\end{document}